\documentclass[aps,prl,showpacs,superscriptaddress,prbib, twocolumn]{revtex4}
\usepackage{graphicx}
\usepackage{float}
\newcommand{\Fig}[1]{Fig.~\ref{#1}}

\begin{document}

\author{Clifford W. Hicks}
\author{Thomas M. Lippman}
\affiliation{Geballe Laboratory for Advanced Materials, Stanford University, Stanford, California, 94305, USA
and Stanford Institute for Materials and Energy Sciences, SLAC National Accelerator Laboratory, 2575 Sand Hill
Road, Menlo Park CA 94025 }
\author{Martin E. Huber}
\affiliation{Departments of Physics and Electrical Engineering, University of Colorado Denver, Denver, Colorado, 80217, USA}
\author{James G. Analytis}
\author{Jiun-Haw Chu}
\author{Ann S. Erickson}
\author{Ian R. Fisher}
\author{Kathryn A. Moler}
\affiliation{Geballe Laboratory for Advanced Materials, Stanford University, Stanford, California, 94305, USA
and Stanford Institute for Materials and Energy Sciences, SLAC National Accelerator Laboratory, 2575 Sand Hill
Road, Menlo Park CA 94025 }


\title{Evidence for Nodal Superconductivity in LaFePO from Scanning SQUID Susceptometry}

\date{2 April 2009}

\maketitle

{\sffamily We measure changes in the penetration depth $\lambda$ of the $T_c \approx 6$~K
superconductor LaFePO. In the process scanning SQUID susceptometry is demonstrated as a technique for
accurately measuring {\it local} temperature-dependent changes in $\lambda$, making it ideal for studying
early or difficult-to-grow materials. $\lambda$ of LaFePO is found to vary linearly with temperature from 0.36
to $\sim$2~K, with a slope of 143$\pm$15 \AA/K, suggesting line nodes in the superconducting order
parameter. The linear dependence up to $\sim T_c/3$ is similar to the cuprate superconductors,
indicating well-developed nodes.}

Research on the iron pnictide superconductors has been intense over the past year. Most attention has focused
on arsenic-based materials, which have the highest transition temperatures, but which only superconduct at
ambient pressure when doped, resulting in intrinsic disorder.
Superconductivity in LaFePO was announced in 2006~\cite{Kamihara}. Most likely it is the fully
stoichiometric compound that superconducts, with $T_c \approx 6$~K; very clean residual resistivities below
0.1 m$\Omega$-cm have been obtained~\cite{Analytis}. How similar LaFePO will prove to be to the higher-$T_c$
arsenide compounds is not clear; although LaFePO does not show the magnetic order found in the arsenide
compounds, its electronic structure has been found to be very similar~\cite{Coldea}.

The temperature dependence of the penetration depth provides information on the 
superconducting order parameter (OP). OPs with line
nodes are known to result in a $T$-linear dependence of $\Delta\lambda(T) \equiv \lambda(T)-\lambda(0)$ at low
$T$~\cite{Annett}. Scattering modifies this dependence to $T^2$~\cite{Hirschfeld}. A
fully-gapped OP results in an exponential dependence $\Delta\lambda \propto
T^{-1/2}\exp(-T_0/T)$~\cite{Tinkham}. For the iron pnictide superconductors, proposed OPs include nodal and
nodeless $s$~\cite{Mazin, Kuroki, Wang, Chen, Chubukov}, $s+d$~\cite{Seo},
$p$~\cite{Lee, Qi} and $d$~\cite{Si, Kuroki, Qi, Yao}. Most of these predictions
are based on calculations in an unfolded, 1 Fe per unit cell Brillouin zone, which neglects the avoided
crossings between the electron pockets in the true zone~\cite{Coldea}. These avoided crossings could significantly
alter the nodal structure of the OP. It is also a possibility that different pnictide superconductors,
although electronically similar, have different OPs~\cite{Scalapino1}.

Radio-frequency tunnel diode resonator and microwave cavity perturbation measurements on iron arsenide
superconductors have shown both power-law and exponential temperature dependences of $\Delta\lambda$.
Power-law dependence has been found in Ba(Fe$_{1-x}$Co$_x$)$_2$As$_2$~\cite{Gordon}, with the exponent $n$
varying between 2.0 and 2.6 with doping. $n \approx 2$ has been found in
Ba$_{1-x}$K$_x$Fe$_2$As$_2$~\cite{MartinBaK}, NdFeAsO$_{0.9}$F$_{0.1}$~\cite{MartinNd} and
LaFeAsO$_{0.9}$F$_{0.1}$~\cite{MartinNd}. Exponential behavior has been found in
Ba$_{1-x}$K$_x$Fe$_2$As$_2$~\cite{Hashimoto_Ba}, 
PrFeAsO$_{1-y}$~\cite{Hashimoto_Pr} and SmFeAsO$_{0.8}$F$_{0.2}$~\cite{Malone}.

In LaFePO, nearly linear dependence of $\Delta\lambda$ on $T$ to below 150~mK has been reported by Fletcher
{\it et al}~\cite{Fletcher}, using an RF tunnel diode circuit.  However early LaFePO samples have had
irregular shapes, which complicate RF and microwave measurements: to isolate $\lambda_{ab}$ the magnetic field
of the excitation must be specifically oriented relative to the crystal axes, and at lower frequencies
knowledge of the sample size is necessary to extract $\Delta \lambda$. Fletcher {\it et al} report these
slopes $d\lambda/dT$ on three samples: 412, 436 and 265 \AA/K (over $0.7<T<1.0$~K). The magnitude of
$d\lambda/dT$ constrains the number and opening angle of nodes, so confirmation with additional
measurement is desirable.

\begin{figure}[ptb]
\includegraphics[width=3.25in]{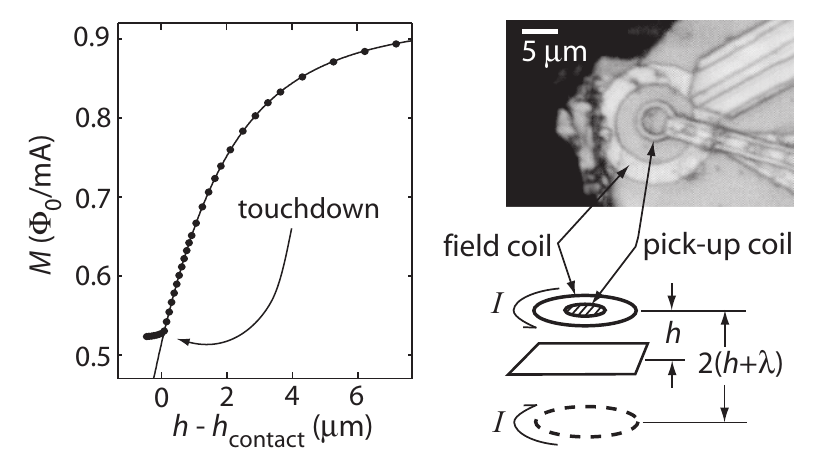} 
\caption{\label{suscVsVz} Left: Field coil--pick-up loop mutual
inductance $M$ against height above the sample; $h_{\mathrm{contact}}$ is the height at which the
corner of the susceptometer chip contacts the sample. The line is a fit of eq.~1. Right: photograph of the
front end of the susceptometer, and a schematic of the susceptometer-sample geometry. The dashed loop is the
image field coil.}
\end{figure}

SQUID susceptometry has been demonstrated as a technique for observing superconducting
transitions~\cite{Gardner}, and has been used to determine the Pearl length $\Lambda$ of thin
superconducting films, for $\Lambda \sim$10--100 $\mu$m~\cite{Tafuri}. We extend this technique to measurement
of nm-scale changes in local $\lambda$ with varying sample temperature. Our susceptometer is a niobium-based
design~\cite{Huber}; its front end is shown in \Fig{suscVsVz}. The pick-up loop is part of a SQUID, and an
excitation current (in this work, at 1071 Hz) is applied to the field coil. What is measured is the field
coil--pick-up loop mutual inductance $M$. The susceptometer chip is polished to a point, aligned at an angle
relative to the sample
(in this work, $\approx 16^\circ$), and mounted onto a 3-axis scanner. The Meissner response of
superconducting samples partially shields the field coil, so $M$ decreases as the susceptometer approaches the
sample.

The schematic in \Fig{suscVsVz} shows a model of the susceptometer. The field coil is taken as a wire loop of
radius $R$ at a height $h$ above the superconductor surface, and the superconductor response field 
as an image coil placed a height $2h_{\mathrm{eff}}$ beneath the field coil, where the effective height
$h_{\mathrm{eff}} = h+\lambda$. The flux through the pick-up coil (radius $a$) is taken as the field at its
center times its area.  All coils are taken parallel to the surface, neglecting the alignment angle. These
approximations give a conversion between $M$ and $h_{\mathrm{eff}}$:
\begin{equation}
M = \frac{\mu_0}{2}\pi a^2\left(\frac{R^2}{\left(R^2+4h_{\mathrm{eff}}^2\right)^{3/2}} - \frac{1}{R}\right).
\end{equation}

To measure changes in $\lambda$ the susceptometer is placed in contact with a flat $ab$-plane area of the
sample, and the sample temperature $T$ is varied. The contact is sufficient to overcome system vibration
but weak enough to avoid excessive thermal coupling (the susceptometer is maintained at $\approx 0.3$~K). The
contact keeps $h$ constant, so changes to $h_{\mathrm{eff}}$ are changes in $\lambda_{ab}$: we
are using the fact that, for $h \gg \lambda$, the response field of the superconductor is a function of
$h_{\mathrm{eff}}$ alone~\cite{Kogan}. In this sense, the physical origin of eq.~1 is irrelevant as long as it
accurately models the dependence of $M$ on $h$, which \Fig{suscVsVz} shows to be the case. $R$ and $a$ are
fitting parameters; they approximately match the actual dimensions of the susceptometer, but with precise
values that vary with alignment angle and sample surface orientation; $R$ and $a$ are obtained separately for
each sample.  Crucial to this measurement, if the susceptometer is over a flat $ab$ surface then the relevant
penetration depth is $\lambda_{ab}$ alone, even with nonzero alignment angle~\cite{Kogan}. Due to the
alignment angle, the minimum $h$ is $h_{\mathrm{contact}}\approx3$ $\mu$m.

What is the accuracy of measurement of $\Delta\lambda$? 
The fit to eq.~1 returns $R$ and $a$ consistent with a particular 
conversion constant, in $\mu$m/V, between $h$ and applied voltage to the scanner, which is measured
separately, in this work with $\pm$5\% accuracy. All $\Delta\lambda$ quoted in this work have this $\pm$5\%
systematic uncertainty. Also, deviations from the fit give errors on $\Delta\lambda$ up to 1.5\%. At large
$\lambda$ the assumption that the response field is a function of $h+\lambda$ breaks down; numerical
simulation shows that, at $h=3$ $\mu$m and $\lambda(0)=5000$ \AA, this assumption leads $\Delta \lambda$ to be
underestimated by 1\% at $\Delta \lambda=5000$ \AA\ and 4\% at 10,000 \AA.
Thermal gradients from the susceptometer-sample contact have minimal effect: control tests
on sapphire show that the change in $h_{\mathrm{eff}}$ attributable to these gradients is no more than
$\sim20$ \AA\ for $T$ varying between 1 and 8~K. By tracking $T_c$ of LaFePO, we determine that contact
locally cools the sample by only $\sim$40~mK at $T=6$~K.

\begin{figure}[ptb]
\includegraphics{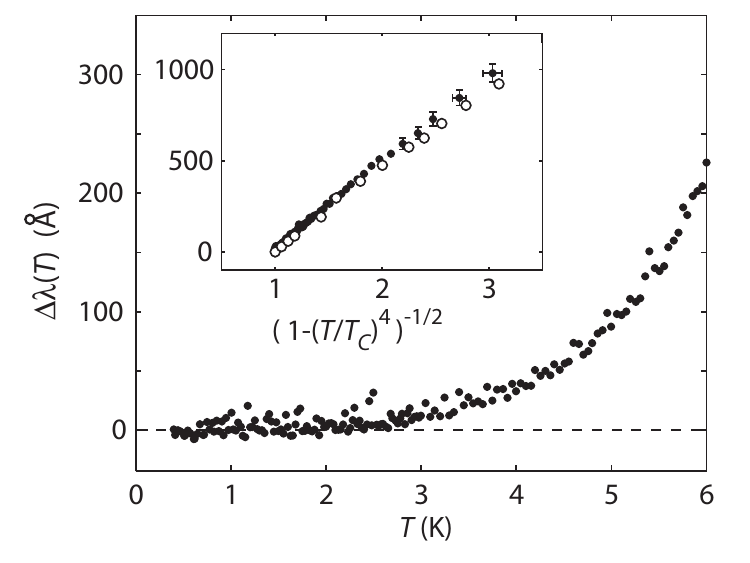}
\caption{\label{leadPenDepth} $\Delta\lambda(T) \equiv \lambda(T)-\lambda(0)$ of Pb. A drift has been
subtracted as described in the text. Inset: Open symbols: measurement of Gasparovic and
McLean~\cite{Gasparovic}. Filled: present data; the vertical error bars are the systematic $\pm5$\%
error on all $\Delta\lambda$ data in this Letter.}
\end{figure}

As a test we measure the penetration depth of a lump of industrial-grade lead; the results and comparison with
an earlier measurement are shown in \Fig{leadPenDepth}. A $\sim100$ \AA-scale downward drift of
$h_{\mathrm{eff}}$, due to the sensor gradually pressing a dent into the soft lead surface, is
subtracted from our data. The drift rate is $T$-independent and was measured separately from the data in
\Fig{leadPenDepth}, so the flatness of $\Delta\lambda$ at low $T$ is real.

\begin{figure}[ptb]
\includegraphics{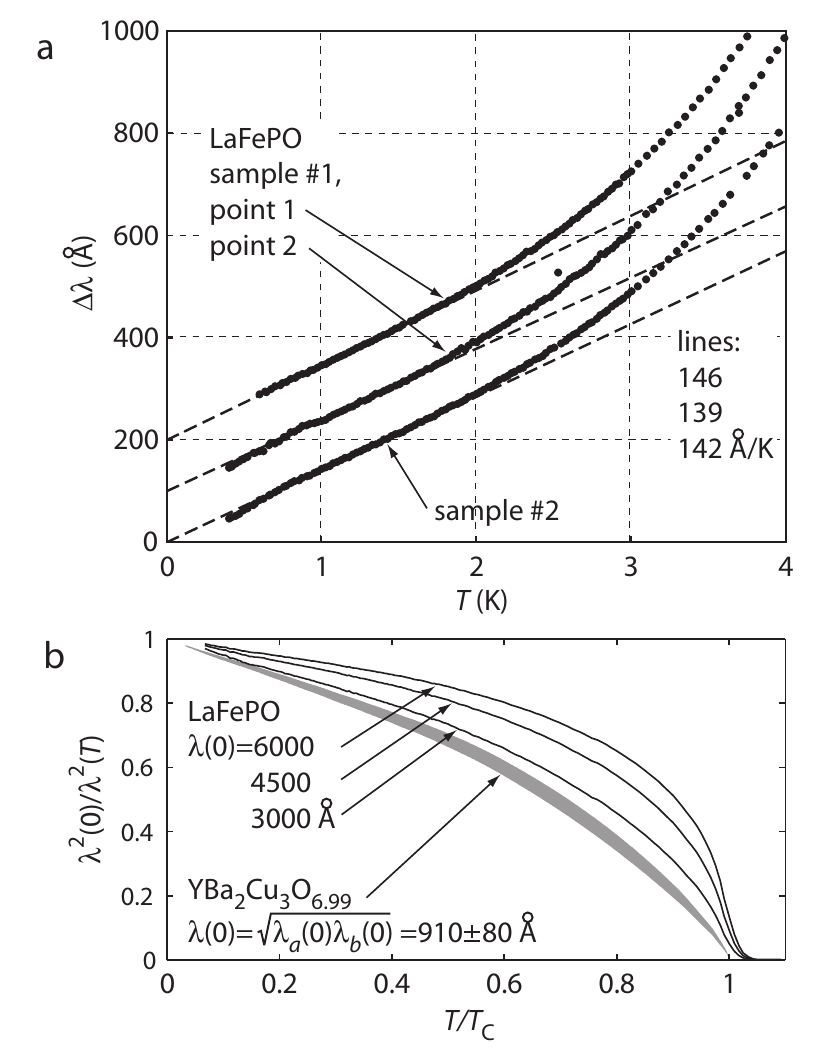}
\caption{\label{penDepth} Top: $\Delta\lambda$ of two LaFePO specimens, at the points indicated in
\Fig{images}. The lines were fit over $0.7<T<1.6$~K.
Bottom: black lines are possible superfluid densities for LaFePO sample~\#1, point~2, with
different $\lambda(0)$. Shaded area: superfluid density of YBa$_2$Cu$_3$O$_{6.99}$
($1/(\lambda_a\lambda_b)$), from~\cite{Bonn} and~\cite{Pereg-Barnea}; the width of the shaded area reflects
uncertainty in $\lambda(0)$.}
\end{figure}

\begin{figure}[ptb]
\includegraphics{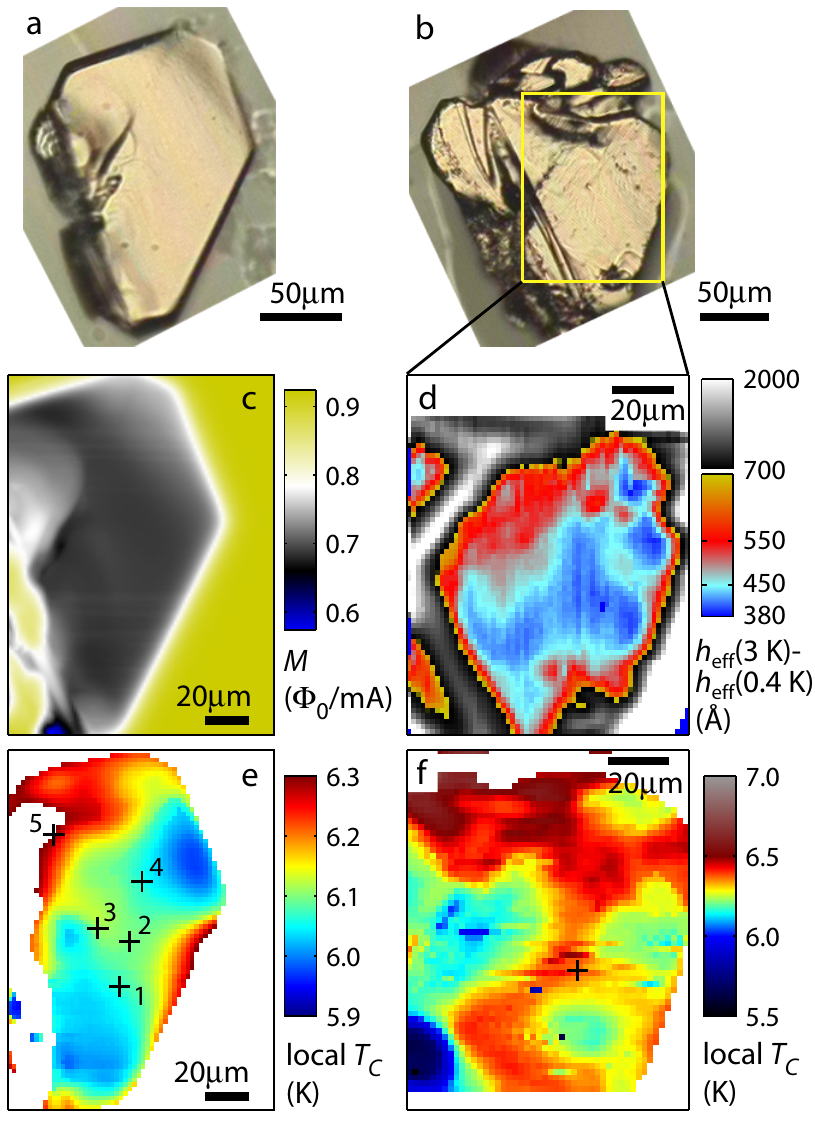}
\caption{\label{images} (a,b) LaFePO specimens \#1 and \#2.  (c) Susceptibility scan of
\#1 at $T=0.4$~K. Over the superconductor $M$ is reduced from its vacuum level by the Meissner response.
(d)~Change in $h_{\mathrm{eff}}= h+\lambda$ between 0.4 and 3~K over specimen \#2. (e,f) Maps of local $T_c$,
over the same areas as in (c) and (d). The crosses indicate the points where $\Delta\lambda(T)$ data were
collected.}
\end{figure}

\Fig{penDepth} shows the main result of this work: $\Delta\lambda_{ab}$ vs. $T$ for two LaFePO
crystals (at the points indicated in \Fig{images}(e) and (f)). For both data sets $\Delta\lambda$ was recorded
over multiple temperature sweeps, both warming and cooling, and found to follow the same path. $\lambda$ is
seen to vary nearly linearly with temperature. Fitting $\Delta\lambda=A+BT^n$ over $0.7 < T < 1.6$~K, from top
to bottom $n$=1.22(4), 1.13(10) and 0.97(5) are obtained for the three curves in
\Fig{penDepth}(a).

Photographs of the two LaFePO specimens are shown in \Fig{images}. An example of a susceptibility scan (a scan
of the spatial variation in $M$) is shown in \Fig{images}(c). Because $M$ varies strongly with $h$, features
in individual scans mainly reflect surface topography.
More useful is comparison of scans at different $T$:
{\it e.g.} \Fig{images}(d) shows a map of $h_{\mathrm{eff}}(3\; {\mathrm K})-h_{\mathrm{eff}}(0.4\; {\mathrm K})$ on sample \#2, which
reveals two useful facts: (1) Where the sample surface is not flat $\lambda_c$ mixes in strongly and $\Delta
h_{\mathrm{eff}}$ is large; one needs to be at least $\sim10$ $\mu$m from an edge to measure $\lambda_{ab}$.
(2) Where the sample is flat, and $\Delta h_{\mathrm{eff}}=\Delta\lambda_{ab}$, $\Delta\lambda_{ab}$
is homogeneous to within $\sim5$\%; areas of moderately increased $\Delta h_{\mathrm{eff}}$ are
areas where the surface is pitted.

Maps of local $T_c$, shown in \Fig{images}(e) and (f), are made by performing susceptibility scans
at various $T$ and extracting $\Delta h_{\mathrm{eff}}(T)$. Most areas of
the samples show weak tails of superfluid density persisting a few 0.1~K beyond the dominant local $T_c$
(and in places beyond 7~K), which give uncertainty to estimates of the dominant $T_c$. Our criterion for
determining local $T_c$ is based on superfluid density: the $\lambda(0)=4500$ \AA\ superfluid density curve in
\Fig{penDepth}(b) is taken as a reference, and in the scans the local $\Delta h_{\mathrm{eff}}$ is taken as
$\Delta\lambda$. The local $\lambda(0)$ (which varies with topography) and $T_c$ are varied to obtain the
best fit to the reference. Varying the reference $\lambda_{ab}(0)$ by 1000 \AA\ varies the calculated $T_c$'s
by $\sim0.1$~K.

$\lambda(0)$ could in principle be extracted from the geometry of the susceptometer and its contact with the
sample surfaces; however the uncertainties are large. From the variation of $M$ with surface orientation and
SEM images of the susceptometer a plausible contact point on the susceptometer can be identified, and
comparison with the lead specimen indicates that $\lambda_{ab}(0)$ of LaFePO likely falls in the range
3500--5500 \AA.

At the five points on sample \#1 indicated in \Fig{images}(e), $d\lambda/dT$ over the linear portion of
$\Delta\lambda$ is 146, 139, 136, 150 and 205 \AA/K, and at the single measurement point on sample \#2, 142
\AA/K. The 205 \AA/K measurement was at a point with significant topography and can be excluded.  Taking into
account the 5\% and 1.5\% uncertainties, $d\lambda/dT$ is 143$\pm$15 \AA/K.

The superfluid densities $\rho_S \equiv 1/\lambda^2$ of LaFePO and YBa$_2$Cu$_3$O$_{6.99}$ are compared in
\Fig{penDepth}. The linear portion of $\rho_S$ persists to a similar fraction of $T_c$ in both
materials, indicating that the nodes in LaFePO must be well-formed, as in YBa$_2$Cu$_3$O$_{6.99}$--- the
magnitude of the gap on either side of the nodes must be similar.
In contrast, accidental nodes in
nodal $s$ orders may result in very asymmetric $+$ and $-$ lobes~\cite{Scalapino2}. Also apparent in
\Fig{penDepth} is that $\rho_S$ of LaFePO rises very sharply on cooling just below $T_c$. If the pairing is
mediated by magnetic fluctuations then such a sharp rise may result from a gapping-out of low-frequency,
pair-breaking fluctuations~\cite{Monthoux}.
\begin{figure}[ptb]
\includegraphics{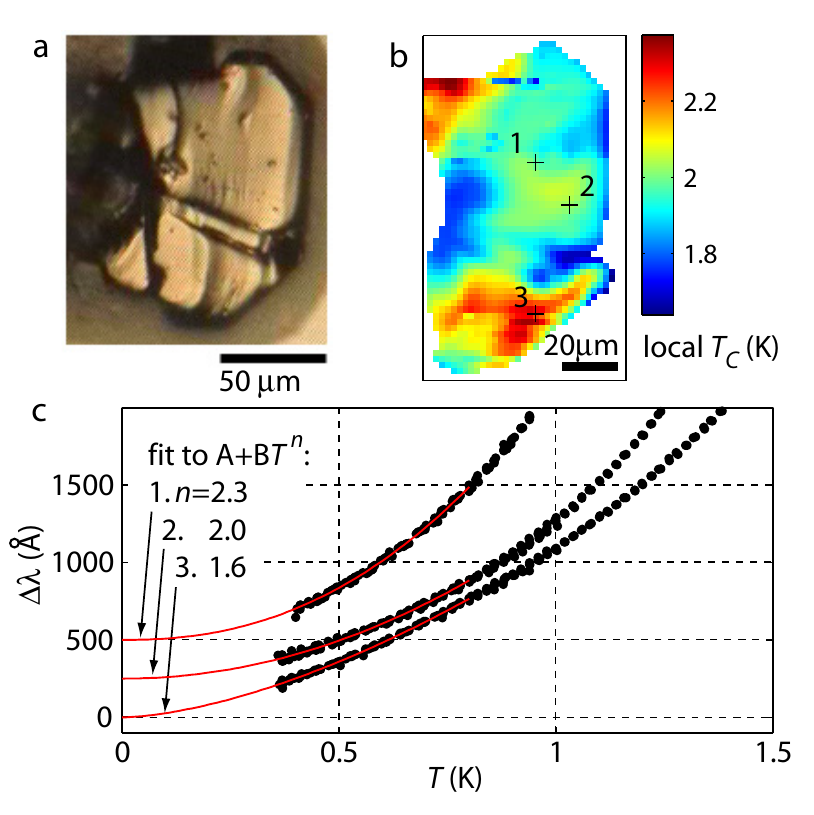}
\caption{\label{2Ksample} (a) Photograph and (b) map of local $T_c$ of a $T_c \approx 2$~K specimen. The
relative error on local $T_c$'s is $\sim 0.1$~K. (c) $\Delta\lambda(T)$ at the three points indicated in (b).
Because the sample surface is not flat, these measured $\Delta \lambda$ include some fraction of $\lambda_c$.}
\end{figure}

An intriguing possibility of highly asymmetric nodal $s$ orders is that scattering might lift the nodes
altogether,
resulting in exponential rather than $T^2$ dependence of $\Delta\lambda$ at
low $T$~\cite{Scalapino2}. In this work we also studied a $T_c \approx 2$~K LaFePO specimen. The reason for the
anomalously low $T_c$ is unclear; electron probe microanalysis shows no impurities (to the $\frac{1}{2}$\%
level), and no anomalies in the La, Fe and P concentrations (to within $\frac{1}{2}$, $\frac{1}{2}$, and 2\%).

\Fig{2Ksample} a shows the 2~K specimen. Compared with the 6~K samples $T_c$ varies more widely both on large
and small length scales: at each point studied strong tails of superfluid density extend well above the
dominant local $T_c$.  $\Delta\lambda$ versus temperature was recorded at three locations.  Fitting to a power
law over $T<0.8$~K, exponents $n$=2.3$\pm$0.1, 2.0$\pm$0.1 and 1.6$\pm$0.1 are obtained.  The dominant
local $T_c$'s at these three locations are 2.1$\pm$0.2, 2.0$\pm$0.1 and 2.5$\pm$0.1~K, respectively;
deviations from $n$=2 may in part reflect variation in the local $T_c$ (or local $T_c$ distribution). On data
up to 0.8~K, power law fits perform better than exponential fits. Within our precision and temperature limits,
$\Delta\lambda(T)$ is consistent with dirty nodal superconductivity, and with many of the measurements on
As-based materials.

In conclusion, we have observed a linear temperature dependence of $\Delta\lambda_{ab}(T)$ below $\sim T_c/3$ in
LaFePO and accurately measured its slope, 143$\pm$15 \AA/K, using a local technique. The large temperature
range of linear $\lambda_{ab}(T)$ indicates well-formed nodes.

This work was funded by the United States Department of Energy (DE-AC02-76SF00515).  We thank Douglas
Scalapino, John Kirtley, David Broun, Vladimir Kogan and Walter Hardy for useful discussions.

\end{document}